\documentstyle[eqsecnum,preprint,tighten,aps]{revtex}
\newfont{\largemi}{cmmi10}
\newfont{\smallmi}{cmmi6}

\begin{document}
\draft

\title {Energy Centroids of Spin $I$ States  by Random Two-body Interactions}
\author{Y. M. Zhao$^{a,b,c}$, A. Arima$^{d}$, and K. Ogawa$^e$}

\vspace{0.2in}
 \address{
  $^a$ Department of Physics, Shanghai Jiao Tong
University,
 200240,   China \\
$^b$ Cyclotron Center,  Institute of Physical Chemical
 Research (RIKEN), \\
Hirosawa 2-1,Wako-shi,  Saitama 351-0198,  Japan  \\
$^c$  Center of Theoretical Nuclear Physics, National Laboratory
of Heavy Ion Accelerator,
Lanzhou 730000, China \\
$^d$ The Science Museum, Japan Science Foundation, 2-1
Kitanomaru-koen,
Chiyodaku, Tokyo 102-0091, Japan \\
$^e$ Department of Physics, Chiba University, Yayoi-cho 1-33,
Inage, Chiba 263-8522, Japan   }

\date{\today}
\maketitle

\begin{abstract}
In this paper we study the behavior of energy centroids (denoted
as $\overline{E_I}$) of spin $I$  states in the presence of random
two-body interactions,
 for    systems  ranging  from  very simple
systems (e.g.   single-$j$ shell for very small $j$) to very
complicated systems (e.g., many-$j$ shells with different parities
and with isospin degree of freedom). Regularities of
$\overline{E_I}$'s  discussed in terms of
 the so-called geometric chaoticity
(or quasi-randomness of two-body coefficients of fractional
parentage)  in earlier works  are found to  hold even for very
simple systems in which   one cannot assume the  geometric
chaoticity.    It is shown  that the inclusion of isospin and
parity does not ``break" the regularities of $\overline{E_I}$'s.
\end{abstract}

\pacs{PACS number:   05.30.Fk, 05.45.-a, 21.60Cs, 24.60.Lz}

\newpage

Low-lying spectra of many-body systems with an  even
 number of particles  were examined
 by Johnson, Bertsch and Dean in Ref. \cite{Johnson}  by
using  random two-body interactions (TBRE). Their
  results showed a preponderance of ${\rm spin}^{\rm parity}=0^+$ ground states.
  Many efforts have been made to understand
this very interesting  observation  and
to study other regularities of many-body
 systems in the presence of random interactions. For instance,
  studies of odd-even staggering of binding energies, generic
collectivity,   behavior of energy centroids for spin $I$ states,
correlations, have  attracted  much attention in recent years. See
Ref. \cite{review1} for a recent review.

Among   many works along the context of regularities under the
TBRE Hamiltonian, regularities of energy centroids (denoted as
$\overline{E_I}$'s) of  spin $I$ states  are very interesting. We
denote by ${\cal P}(I)$   the probability that $\overline{E_I}$ is
the lowest energy among all $\overline{E_{I'}}$'s.
 It was shown in Refs. \cite{Arima,Zhaox} that ${\cal P}(I)$
 is large only when  $I\simeq I_{\rm min}$ or $I \simeq I_{\rm max}$.
One thus divides the TBRE into two subsets, one subset  has
  $\overline{E_{I \simeq I_{\rm min}}}$ as the lowest energy,
  and the other  has  $\overline{E_{I  \simeq I_{\rm max}}}$ as
 the  lowest energy.
We define  $\langle \overline{E_I} \rangle_{\rm min}$
 ($\langle \overline{E_I} \rangle_{\rm max}$) as the value
  obtained by averaging
 $\overline{E_I}$ over the subset where
 $\overline{ E_{I \simeq I_{\rm min}}}$
 ($\overline{ E_{I \simeq I_{\rm max}}}$) is the lowest energy.
It was  shown  in Ref. \cite{Zhaox}  that
 $\langle \overline{E_I} \rangle_{\rm min} \simeq
CI(I+1)$ and $\langle \overline{E_I} \rangle_{\rm max} \simeq
C\left[ I_{\rm max} (I_{\rm max}+1) - I(I+1) \right]$, where
$C$ is a constant depending on the occupied single-particle orbits
 and the choice of the ensemble.
 These features were discussed  by using  the quasi-randomness of
two-body coefficients of fractional parentage   (cfp's) in Ref.
\cite{Zhaox},  and  by using other statistical views  in Refs.
\cite{Mulhall1,Kota}.

The purpose of this Brief Report is to revisit regularities of
$\overline{E_I}$. We shall show that the above regularities  of
${\cal P}(I)$'s and $\langle \overline{E_I} \rangle_{\rm min}$'s
hold even  for  very simple systems in which  one cannot assume
that two-body cfp's are random.  Previous studies of
$\overline{E_I}$ under random two-body interactions have been
restricted to identical fermions in one-$j$ shell or two-$j$
shells.  Here we shall extend the study of $\overline{E_I}$ under
random interactions  to systems of many-$j$ shells with the
inclusion of parity  and/or isospin.

In this paper we use  the general shell model Hamiltonians defined
in Ref. \cite{review1}, and take the TBRE of Ref. \cite{Johnson}
for   two-body matrix elements. $\overline{E_I}$  and ${\cal
P}(I)$  for $``\pm"$  parity states are denoted by
$\overline{E_{I^{\pm}}}$  and ${\cal P}(I^{\pm})$, respectively.
The number of particles is denoted by $n$. Proton (neutron) degree
of freedom is denoted by ``${\rm p}$" (``${\rm n}$"). Our
statistics are based on 1000 sets of the TBRE Hamiltonian.

We begin with simple systems, i.e., fermions in a   small
single-$j$ shell ($j\le \frac{7}{2}$) and bosons with a small spin
$l$. First we study fermions in a $j=5/2$ or $7/2$ shell.
$\overline{E_{I}}$  for three fermions were given in Eqs. ~(2.1)
and (D1) of Ref. \cite{Zhao-1}. $\overline{E_{I}}$ for four
fermions in a $j=7/2$ shell can be obtained based on Eq. (5) of
Ref. \cite{PRC-2001}.  ${\cal P}(I)$'s obtained by using the TBRE
Hamiltonian and those by applying the empirical rule of Ref.
\cite{Zhao-1} are plotted and compared in Fig. 1(a-c),  where a
reasonable agreement is achieved. One sees that ${\cal P}(I)$'s
are large for $I\simeq I_{\rm min}$ or $I\simeq I_{\rm max}$,
except that this pattern is not very striking for Fig. 1(a) where
there are only three $\overline{E_{I}}$'s given by three two-body
matrix elements. For $j=\frac{7}{2}$, ${\cal P}(I)$'s are  small
for ``medium" $I$.

Let us look at  $\langle \overline{E_{I}} \rangle_{\rm min}$'s,
which are obtained by averaging  $\overline{E_I}$ over the subset
with $\overline{E_{I \simeq I_{\rm min}}}$ being the lowest
energy. We plot $\langle \overline{E_{I}} \rangle_{\rm min}$'s for
$n=3$ with $j=5/2$, $n=3$ with $j=7/2$, and $n=4$ with $j=7/2$ in
Fig. 1 (a$'$-c$'$), respectively. We see that $\langle
\overline{E_{I}} \rangle_{\rm min}$'s are approximately
proportional to  $I(I+1)$.

Now we study bosons with small spin $l$. The case with  $l=1$ ($p$
bosons) can be easily understood:  There is only one state for
each $I$; ${\cal P}(I)$=$50\%$ for $I=I_{\rm min}$ or $I=I_{\rm
max}$,  and zero for other $I$ values; $\overline{E_I}$ follows
the $\overline{E_I}=C I(I+1)$ relation precisely.

As for $l=2$ ($d$) bosons, we  predict  ${\cal P}(I)$ values for
$n=3$, 4, 5, and  6 by applying the empirical rule of Ref.
\cite{Zhao-1} and compare them with those obtained by using the
TBRE Hamiltonian in Fig. 2(a-d).    Fig. 2(a$'$-d$'$) plots
$\langle \overline{E_{I}} \rangle_{\rm min}$ versus $I(I+1)$. A
linear correlation between these two quantities is easily noticed.
Because all eigenvalues of $d$-boson systems are known, one can
study ${\cal P}(I)$ and correlation between $\langle
\overline{E_I} \rangle_{\rm min} $ and $I(I+1)$ at a more
sophisticated level. From  Eqs. (2.7)-(2.8) of Ref. \cite{Zhao-1},
we have
\begin{equation}
\overline{E_I} = E'(n) + \frac{1}{70} (10 c_2 - 7c_0 - 3c_4)
\overline{v (v+3)} + \frac{1}{14} (c_4 - c_2) I(I+1) ~. \label{d1}
\end{equation}
Eq. (\ref{d1}) shows that there are three terms in
$\overline{E_I}$'s: the first is just a constant and the second is
related to $\overline{v(v+3)}$, the difference of which between
neighboring $I$ is large  for low $I$ and is negligible  for $I\gg
I_{\rm min}$;  the third one is $I(I+1)$, which is small for low
$I$ and becomes dominant for large $I$. Thus  ${\cal P}(I)$ is
sensitive to the value of $\overline{v(v+3)}$ only for low $I$.
Let us evaluate $C$ in the formula  $\langle \overline{E_I}
\rangle_{\rm max} \simeq C\left[ I_{\rm max} (I_{\rm max}+1) -
I(I+1) \right]$ for $d$-boson systems.   A simple assumption here
is $(c_4 - c_2) \le 0$. The  $C$ value can  be evaluated  by
averaging $\frac{1}{14} (c_4 - c_2)$ under such a requirement:
\begin{eqnarray}
&& C= \frac{1}{14}\times  2 \times \frac{1}{\sqrt{2 \pi}}
\frac{1}{ \sqrt{2}} \int_0^{-\infty} x {\rm e}^{-\frac{x^2}{4}}
{\rm d}x = -\frac{1}{7 \sqrt{\pi}}
 \simeq - 0.0806 ~.  \label{C-d}
\end{eqnarray}
The value of $C$ obtained  by using the TBRE Hamiltonian is $\sim
-0.73$ if $\overline{E_{I\simeq I_{\rm max}}}$ is the lowest
energy and  $\sim 0.070$ if $\overline{E_{I\simeq I_{\rm min}}}$
is the lowest energy. The difference between our predicted $|C|$
of Eq. (\ref{C-d}) and those obtained by the TBRE Hamiltonian
comes from the complexity of $\overline{v(v+3)}$, which can be
formulated analytically for any $I$.

Now we come to systems of many-$j$ shells with the inclusion of
parity. Let us exemplify by a system with four identical nucleons
in two-$j$ shells: $j_1^{\pi}=5/2^+$ and $j_2^{\pi}=3/2^-$.
 Fig. 3 plots our calculated ${\cal P}(I^{\pm})$ in terms of $I^{\pm}$ and
 $\langle  \bar{E}_{I^{\pm}} \rangle_{\rm min}$ in terms of
$I^{\pm}(I^{\pm}+1)$.  One sees that both ${\cal P}(I^{\pm})$  and
$\langle \bar{E}_{I^{\pm}} \rangle_{\rm min}$  behave similarly as
${\cal P}(I)$ and  $ \langle \overline{E_I} \rangle_{\rm min}$ in
Ref. \cite{Zhao-1}.  The predicted ${\cal P}(I^{\pm})$'s  by using
the empirical rule of Ref. \cite{Zhao-1} are reasonably consistent
with those obtained numerically by using the TBRE Hamiltonian.
According to  our statistics,  $\sum_{I^+} {\cal P}(I^+) = 41.3\%$
while $\sum_{I^-} {\cal P}(I^-) = 58.7\%$;
 $C^+=0.0372 \pm 0.0017$ and $C^-=0.0359 \pm 0.0029$.  These
 $C^{\pm}$ values are close to $C$ values obtained for  $d_{3/2}
 d_{5/2}$  shells.  For four identical nucleons in $d_{3/2} d_{5/2}$ shells,
$C=0.0401\pm 0.0017  $.  This suggests that the relation $\langle
\bar{E}_{I^{\pm}} \rangle_{\rm min} \simeq C^{\pm}
I^{\pm}(I^{\pm}+1)$ and the value of $C$ are not sensitive to
parity of  single-particle levels in the model space.

We next study many systems  with the inclusion of isospin degree
of freedom. Fig. 4  presents a few typical examples of ${\cal
P}(I)$ and $\langle \overline{E_{I}} \rangle_{\rm min}$. We see
that ${\cal P}(I)$ is large only when $I\simeq I_{\rm min}$ or
$I\simeq I_{\rm max}$, and that $\langle \overline{E_{I}}
\rangle_{\rm min} \simeq   C I(I+1)$. According to our
calculations,
  $C = 0.0354 \pm 0.0003, 0.0341 \pm 0.0001, 0.0350 \pm 0.0001,
0.0341 \pm 0.0002$ for
 ($n_{\rm p}$, $n_{\rm n}$)=(4,4), (4,5), (4,6) and (6,6) in
  $s_{1/2}d_{3/2}d_{5/2}$ shells.
$C=0.0331 \pm 0.0002, 0.0331 \pm 0.0001, 0.0338 \pm 0.0001$ for
($n_{\rm p}$, $n_{\rm n}$)=(2,4), (4,3), and (4,4) in fictitious
two-$j$ shells $d_{3/2}d_{5/2}$.   $C$ values are very close to
those given by the empirical formula $C \simeq \frac{1}{4\sum_i
j_i^2}$ suggested on the basis of numerical experiments in Ref.
\cite{Zhaox}.

We have also studied systems which include both parity and
isospin. Our results present similar regularities, and suggest
that   $C^{\pm}$ values are   sensitive to  $j$ shells but  not
sensitive to particle numbers,  nor to the isospin degree of
freedom.

Now we discuss the formula of $\overline{E_I}$ obtained in Refs.
\cite{Mulhall1,Kota} by  evaluating the $C$ value in $\langle
\overline{E_I} \rangle_{\rm min} \simeq CI(I+1)$. The coefficient
of the third term in Eq. (9) of Ref. \cite{Kota} is a Gaussian
random number with   width
\begin{eqnarray}
&&\sigma  = \sqrt{ \sum_{{\rm even} ~ J} \left( (2J+1)
 \frac{3 (J(J+1) -  2j(j+1))}{2j^2 (j+1) (2j+1)} \right)^2 } \nonumber \\
&& = \sqrt{
\frac{3(1408 j^6 + 864 j^5 -4296 j^4 -512 j^3 + 5688 j^2 +558 j -945)}{560
j^4 (j+1)^4 (2j+1)^3 } }~.   \label{Kota-0}
\end{eqnarray}
One sees that $\sigma \propto  j^{-5/2}$ when $j \rightarrow
\infty$. The coefficient $C$ in  $\langle \overline{E_I}
\rangle_{\rm min} = CI(I+1)$ is given by   $\sigma  \times
\sqrt{2/\pi}$. The $C$ value based on Refs. \cite{Mulhall1,Kota}
is therefore proportional to $\frac{1}{\sqrt{j^5}}$ at the large
$j$ limit.  This is different from the empirical formula $C \simeq
\frac{1}{4j^2}$. In Table I we list a few $C$ values obtained by
different methods. This table shows that there is a systematic
discrepancy between the predicted $C= \sqrt{2/\pi} \sigma $ with
$\sigma$ given by Eq. (\ref{Kota-0}) and that obtained by using
the TBRE Hamiltonian. The reason for this discrepancy should be
clarified in the future.

To summarize, in this paper we have studied energy centroids of
spin $I$ states in the presence of random two-body interactions.
First, we have shown that the regularities--- ${\cal P}(I)$'s are
large only when $I\simeq I_{\rm min}$ or $I\simeq I_{\rm max}$,
and $\langle \overline{E_{I}} \rangle_{\rm min} \simeq
I(I+1)$---hold approximately even for very simple systems in which
cfp's can not be assumed ``random". These simple systems include
 fermions in a $j=5/2$ or $j=7/2$ shell,
$l=1$ ($p$) bosons, and $l=2$ ($d$) bosons.
We point out that,
although the above regularities of energy centroids of spin $I$ states
are noticed and argued in Refs. \cite{Zhaox,Mulhall1,Kota},
 a  {\it sound} understanding of $\overline{E_{I}}$ is not yet available.
The arguments of Refs. \cite{Zhaox,Mulhall1,Kota} might be part of
the story, and the randomness of  cfp's is not the unique origin
of these regularities.

Second, we have  shown that the above regularities are also robust
 with the inclusion of  parity and/or isospin:  ${\cal P}(I^{\pm})$'s are
 large only when   $I^{\pm} \simeq I_{\rm min}$ or  $I^{\pm}
 \simeq I_{\rm max}$, and $\langle \overline{E_{I^{\pm}}}
 \rangle_{\rm min} \simeq C^{\pm} I^{\pm}(I^{\pm}+1)$. $C^{\pm}$ is
 not sensitive to parity or isospin but sensitive to the value of
 $j$. We note without details that this pattern  also holds for
two-body random interactions which are uniformly distributed.

Finally, we would like to mention two works on the energy
centroids and other trace quantities such as spectral variances.
In Ref. \cite{Zuker} Velazquez and Zuker used  energy centroids
and spectral variances to obtain the lower bound of  energy for
spin $I$ states in the presence of random two-body interactions.
In Ref. \cite{Papenbrock} Papenbrock and Weidenmueller  derived
the distribution of and the correlation between spectral variances
of different spin $I$ states, and discovered a correlation between
spin $I$ ground state probability and its spectral variance
multiplied by a scaling factor. These studies are very
interesting, and further studies along this line are called for.

Acknowledgement: We would like to thank Drs. V. K. B. Kota and W.
Bentz for  discussions and  communications.

\newpage

Captions:

Fig. ~ 1   ~~~    ${\cal P}(I)$'s
 and  $\langle \bar{E}_I \rangle_{\rm min}$'s
 for three fermions in a $j=5/2$ or $7/2$ shell,
and for four fermions in a $j=7/2$ shell.
 $\langle \overline{E_I} \rangle_{\rm min}$'s are obtained by
averaging over the subset of $\overline{E_{I \simeq I_{\rm
min}}}$. One sees that ${\cal P}(I)$'s are large for $I\simeq
I_{\rm min}$ and $I\simeq I_{\rm max}$, except the case  with
$n=3$ and $j=5/2$ for which three states given by  three random
two-body interactions.  A good agreement between ${\cal P}(I)$
obtained by using the TBRE Hamiltonian and that by the empirical
rule of Ref. \cite{Zhao-1} is easily seen. The dotted lines in
(a$'$-c$'$) are plotted to guide the eyes.

\vspace{0.4in}

Fig. ~ 2   ~~~    ${\cal P}(I)$'s  and   $\langle \bar{E}_I
\rangle_{\rm min}$'s for $d$ bosons with $n=3$, 4, 5, and 6.
${\cal P}(I)$'s are large when  $I\simeq I_{\rm min}$ or $I\simeq
I_{\rm max}$. ${\cal P}(I)$'s predicted by using the empirical
rule are reasonably consistent with those obtained by using the
TBRE Hamiltonian. $\langle \bar{E}_I \rangle_{\rm min}$'s  of
these systems are proportional to $I(I+1)$ with fluctuations.

\vspace{0.4in}

Fig. ~ 3   ~~~ ${\cal P}(I^{\pm})$'s  and $\langle
\bar{E}_{I^{\pm}} \rangle_{\rm min}$'s  for four fermions in a
two-$j$ shell: $\frac{5}{2}^+, \frac{3}{2}^{-}$. ${\cal
P}(I^{\pm})$'s for both negative parity and positive parity are
large for $I\simeq I_{\rm min}$ or $I_{\rm max}$. The linear
correlation between $\langle \bar{E}_{I^{\pm}} \rangle_{\rm min}$
and $I(I+1)$ holds for both positive and negative parity states of
this simple system, with  $C^{\pm}$  very close to each other.

\vspace{0.4in}

Fig. ~ 4 ~~~     ${\cal P}(I)$ and   $\langle \bar{E}_I
\rangle_{\rm min}$ for a few systems with both valence neutrons
and protons.

\newpage

{TABLE I. The coefficients $C$ for a single-$j$ shell. The column
``TBRE" is obtained by 1000 runs of the TBRE Hamiltonian. The
column ``empirical" is obtained by the simple formula
$\frac{1}{4j^2}$ suggested in Ref. \cite{Zhaox}. The column ``Z-K"
is obtained by the formulas given in Refs. \cite{Mulhall1,Kota}.
Because $C$ is not sensitive to particle number $n$, all results
are given by $n=4$. It is noted that $C$ values obtained by the
formula of  Refs. \cite{Mulhall1,Kota} is systematically smaller
(about 70-80$\%$) than the ``experimental" values for a single-$j$
shell. }

\vspace{0.3in}

\begin{tabular}{cccc} \hline  \hline
$2j$ & TBRE &  empirical &  Z-K   \\  \hline
 9   &  0.01374  & 0.01235  & 0.01026    \\
 11  &  0.00826  &  0.00826  &  0.006907   \\
 15  & 0.00474 & 0.004444   &  0.003604   \\
 21  & 0.00231 & 0.002268   &  0.001712   \\
 27  & 0.00131 & 0.001372   &  0.000963   \\
  \hline  \hline
\end{tabular}

\vspace{0.3in}

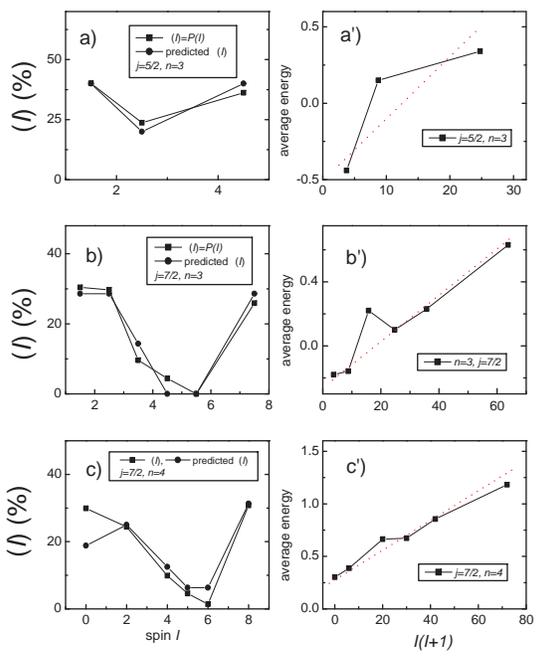
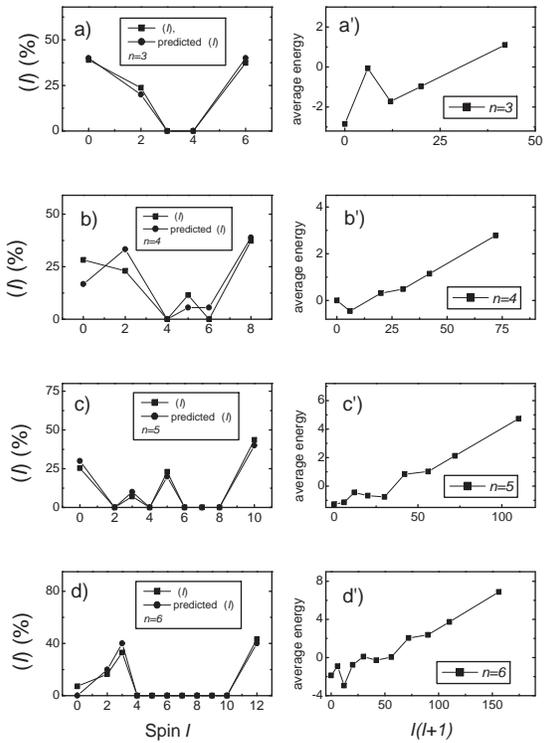
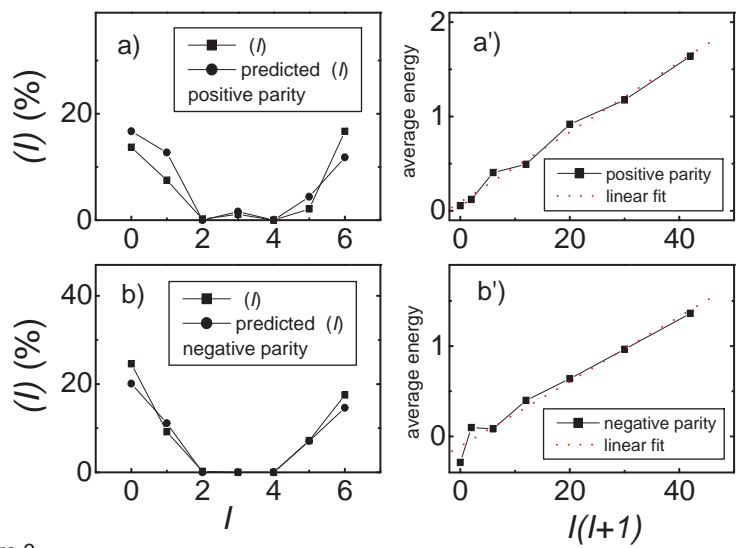
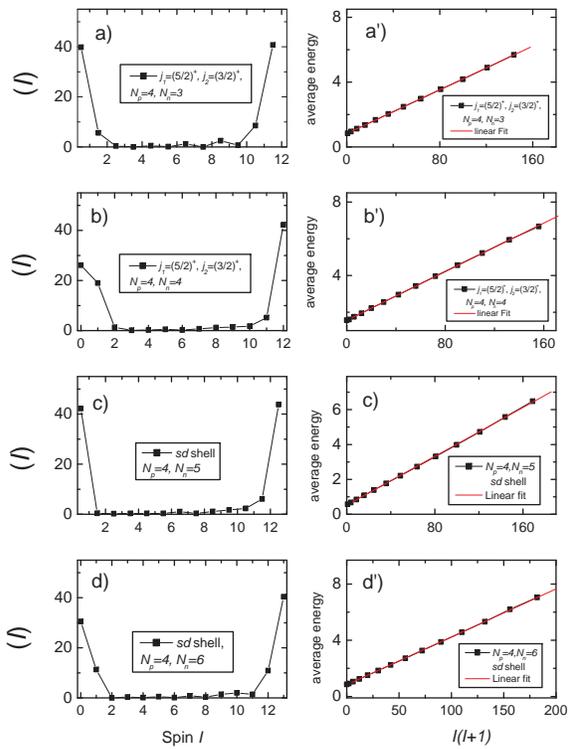
\end{document}